\documentclass[doublecol]{epl2} 
\usepackage{graphicx}
\usepackage{epsfig,bm}
\usepackage{color}

\begin{document}

\title{Resonances in Mie scattering by an inhomogeneous atomic cloud}
\shorttitle{Resonances in Mie scattering by an inhomogeneous atomic cloud}

\author{R. Bachelard\inst{1,2} \and Ph. W. Courteille\inst{2} \and R. Kaiser\inst{3} \and N. Piovella\inst{4}}
\shortauthor{R. Bachelard \etal}

\institute{                    
  \inst{1} University of Nova Gorica, School of Applied Sciences, Vipavska 11c SI-5270 Ajdovšcina, Slovenia\\
  \inst{2} Instituto de F\'{i}sica de S\~{a}o Carlos, Universidade de S\~{a}o Paulo, 13560-970 S\~{a}o Carlos, SP, Brazil\\
  \inst{3} Institut Non Lin\'eaire de Nice, CNRS, Universit\'e de Nice Sophia-Antipolis, 06560 Valbonne, France\\
  \inst{4} Dipartimento di Fisica, Universit\`a Degli Studi di Milano, Via Celoria 16, I-20133 Milano, Italy
}
\pacs{42.50.Ct}{Quantum description of interaction of light and matter; related experiments}
\pacs{42.25.Fx}{Diffraction and scattering}
\pacs{42.25.Gy}{Edge and boundary effects; reflection and refraction}

\abstract{
Despite the quantum nature of the process, collective
scattering by dense cold samples of two-level atoms can be
interpreted classically describing the sample as a macroscopic
object with a complex refractive index. We demonstrate that
resonances in Mie theory can be easily observable in the cooperative scattering
by tuning the frequency of the incident laser field or the atomic
number. The solution of the scattering problem is obtained for
spherical atomic clouds who have the parabolic density characteristic of BECs, and the cooperative radiation pressure
force calculated exhibits resonances in the cloud displacement for dense clouds.
At odds from uniform clouds which show a complex structure including narrow peaks,
these densities show resonances, yet only under the form of quite regular and contrasted oscillations.
}

\maketitle


\maketitle

\section{Introduction}

Mie theory is the well-known solution of Maxwell's equations for
the scattering of electromagnetic radiation by spherical objects
\cite{Mie1908,Lorenz1890}. Via calculation of the electric and magnetic
fields inside and outside the object the theory predicts
the total optical cross section, which determines the amount
of scattered light, and the form factor, which characterizes the
far-field radiation pattern \cite{Hulst,Bohren}. Simple solutions
are available in regimes where the object size $R$ differs very
much from the radiation wavelength $\lambda$, or when the refractive index $m$ is close to unity. For example, for small phase-shifts
$|m-1| R/\lambda \ll 1$ in optically dilute media $|m-1|\ll 1$, one enters the
Rayleigh-Debye-Gans regime whereas for small particles and small phase-shifts, one obtains
Rayleigh scattering by point-like objects.

For objects whose size is of the order of the radiation wavelength
(e.g. water droplets in the atmosphere or in emulsions), Mie's
full theory has to be used to find the scattering pattern. Mie scattering differs from Rayleigh
scattering in several respects. While the intensity of
Rayleigh-scattered radiation scales with the object size as $R^6$
and is identical in forward and backward direction, the intensity
of Mie-scattered radiation is roughly independent of wavelength,
and it is larger in forward than in backward direction. The
greater the particle size, the more light is Mie-scattered
into forward direction. The hallmark of Mie scattering, however,
are the Mie resonances: those are sets of parameters (size,
refraction index, wavelength), where Mie scattering is
particularly strong or particularly weak. The sharpness of some Mie
resonances make them useful for measuring unknown parameters such as particles' size.

Recently, a series of papers demonstrated how collectivities of
point-like Rayleigh-scattering particles may cooperate
\cite{Scully06,Friedberg08,Courteille10,Svindzinsky10,Bienaime10,Bachelard11}
in scattering radiation into the forward direction and the
relationship to Mie scattering was pointed out \cite{Sokolov09,Bender10}.
Here, we show that the theory of collective scattering by smooth
distributions of point-like scatterers is equivalent to Mie
scattering by demonstrating that the premisses of both models are
identical. Hence, we may apply the Mie scattering technique to atomic
clouds as long as their granularity, as
well as collisions and nonresonant atomic interactions, can be
neglected.

In Mie theory, boundary conditions of the scattering object
assume a fundamental role: these are generally sharp since dielectric spheres
typically have homogeneous densities, while atomic clouds, in
general, have parabolic or Gaussian density distributions and smooth
boundaries. Within the framework of the Mie theory, we compare
calculations for homogeneous and parabolic densities,
 and identify the impact of sharp or smooth
boundaries on the occurrence and shape of Mie resonances. Our study reveal that some resonances will persist, although not the sharpest: this means that a jump in the refractive index is not required, as one could expect from classical cavities where smooth gradients of index also allow for mode propagation and selection (e.g. graded-index fibers).

Finally, we predict experimental regimes where resonances in Mie scattering can
be observed in atomic clouds by monitoring the radiation pressure force
acting on the cloud's center-of-mass. To this
aim, we calculate the resonances as a function of atom number
and pump laser detuning. A major advantage of using resonant atoms
as scatters is that the refraction index can be varied over very
large ranges by changing the cloud's density and volume, or simply
by tuning the light frequency. This greatly facilitates the
detection of the resonances.

In this Letter, we first show the equivalence of the cooperative scattering by
two-level atoms and the Mie scattering, assuming a scalar photon
model and reducing the scattering equation for excitation
probability amplitude to a differential Helmholtz equation with
complex refraction index.  Then, we solve the Mie problem and
calculate the radiation pressure force, before discussing the presence of Mie resonances in spherical clouds with parabolic densities. Finally, we draw our conclusions.

\section{From collective atomic scattering to Mie scattering\label{sec:CSWE}}

In his seminal paper\cite{Mie1908}, Gustav Mie proposed an
analytical solution to the scattering of light over extended
objects under the form of infinite series. This theory was
originally developed for homogeneous media of refractive index $m_0$, in which the
elementary solutions to the wave equation are known to be of the
form $\psi_{nl}=j_n(m_0 k_0r)Y_{nl}(\theta,\varphi)$ (where $j_n$ is the spherical Bessel function, and $Y_{nl}$ the spherical harmonics),
while outside they simply read $j_n(k_0r)Y_{nl}(\theta,\varphi)$. The continuity of
the fields at the boundaries eventually determines the scattering
coefficient for each $(n,l)$ mode.

On the other hand, the cooperative scattering  by a sample of $N$
two-level atoms (with random position $\mathbf{r}_j$, transition
frequency $\omega_a$ and linewidth
$\Gamma=d^2\omega_a^3/2\pi\hbar\epsilon_0 c^3$, where $d$ is
the electric dipole matrix element), illuminated by a resonant
uniform field is described by the following coupled
equations~\cite{Courteille10,Svindzinsky10,Bachelard11}
\begin{eqnarray}\label{eqbetaj}
\frac{d\beta_j}{dt}&=&\left(i\Delta_0-\frac{\Gamma}{2}\right)\beta_j-
\frac{i}{2}\frac{d}{\hbar} E_i(\mathbf{r}_j)
\\ &&-\frac{\Gamma}{2}\sum_{m\neq j}\frac{\exp(ik_0|\mathbf{r}_j-\mathbf{r}_m|)}{ik_0|\mathbf{r}_j-\mathbf{r}_m|}\beta_m,\nonumber
\end{eqnarray}
where $j=1,\dots,N$ and $\beta_j$ is the probability
amplitude of excitation of the  $j$th atom, $E_i(\mathbf{r})=E_0
e^{i\mathbf{k}_0\cdot\mathbf{r}}$ is the electric field of the
incident laser and $\Delta_0=\omega_0-\omega_a$  its detuning
with respect to the atomic transition, where $\omega_0=ck_0$. In this approach, short-range dipole terms and polarization effects are neglected~\cite{Friedberg73}.
Neglecting granularity effects, the cloud can be described by a continuous field
$\beta(\mathbf{r},t)$, whose steady-state regime is given by
\begin{eqnarray}
e^{i\mathbf{k}_0\cdot\mathbf{r}} &=&
\left(2\delta+i\right)\tilde\beta(\mathbf{r})\label{eq:gen}
\\ && +\int \mbox{d}\mathbf{r}' \rho(\mathbf{r}')\frac{\exp(ik_0|\mathbf{r}-\mathbf{r}'|)}{k_0|\mathbf{r}-\mathbf{r}'|}
\tilde\beta(\mathbf{r}'),\nonumber
\end{eqnarray}
where $\rho(\mathbf{r})$
is the atomic density, $\delta=\Delta_0/\Gamma$ and we have set
\begin{equation}
\beta(\mathbf{r})=\frac{d E_0}{\hbar\Gamma}\tilde\beta(\mathbf{r}).
\end{equation}
Let us remark that the kernel of
Eq.(\ref{eq:gen}) is the Green function for the Helmholtz equation, that is
\begin{equation}
(\nabla^2+k_0^2)\frac{\exp(ik_0|\mathbf{r}-\mathbf{r}'|)}{|\mathbf{r}-\mathbf{r}'|}
=-4\pi\delta(\mathbf{r}-\mathbf{r}')
\end{equation}
and that $(\nabla^2+k_0^2)\exp(i\mathbf{k}_0\cdot\mathbf{r})=0$. Then, applying
$(\nabla^2+k_0^2)$ on Eq.(\ref{eq:gen}), we obtain that $\tilde\beta(\mathbf{r})$
satisfies the Helmholtz equation \cite{Svindzinsky:cylin,Prasad11}
\begin{equation}
[\nabla^2 +k_0^2 m^2(\mathbf{r})]\tilde\beta(\mathbf{r})=0,\label{eq:web}
\end{equation}
where $m(\mathbf{r})$, the cloud refractive index, is given by
\begin{equation}
m^2(\mathbf{r})=1- \frac{4\pi \rho(\mathbf{r})}{k_0^3\left(2\delta+i\right)}.\label{eq:nr}
\end{equation}
Hence, the cloud of cold atoms acts on the light as a ``classical'' medium of index $m(\mathbf{r})$, whose
imaginary part originates in the single-atom decay term~\cite{note:Prasad}: 
it is here responsible for the absorbing nature of the cloud, and vanishes only in the limit of far-detuned incident laser.

For a cloud with spherical symmetry $m(r)$, the solutions of the wave equation can be decomposed along the orthogonal
basis of the spherical harmonics~\cite{Bohren} as $\sum_{n=0}^\infty\sum_{s=-n}^{n}u_n(r)Y_{ns}(\theta,\varphi)$ where the radial modes $u_n$ satisfy
\begin{equation}
u_n''(r)+2\frac{u_n'(r)}{r}+\left[m^2(r)-\frac{n(n+1)}{r^2}\right]u_n=0.\label{eq:radeq}
\end{equation}
We focus on the axi-symmetric problem, where only the $s=0$ modes are relevant, and the solution write
\begin{equation}
\tilde\beta(\mathbf{r})=\sqrt{4\pi}\sum_{n=0}^\infty \sqrt{2n+1}i^n\beta_n u_n(r)Y_{n0}(\hat{r}).\label{eq:sbn}
\end{equation}
where $\hat r$ is a unit vector in the direction of $\mathbf{r}$. As for the incident wave, it decomposes as
\begin{equation}
\exp(i\mathbf{k}_0\cdot\mathbf{r})= \sqrt{4\pi}\sum_{n=0}^\infty \sqrt{2n+1} i^n j_n(k_0r) Y_{n0}(\hat{r}).
\end{equation}
Since the exponential kernel is diagonal in the basis of the spherical harmonics
\begin{eqnarray}
\frac{\exp(ik_0|\mathbf{r}-\mathbf{r}'|)}{k_0|\mathbf{r}-\mathbf{r}'|}=4\pi i\sum_{n=0}^\infty\sum_{s=-n}^n Y_{ns}(\hat{r})Y_{ns}^*(\hat{r}')\nonumber
\\ \times\left \{
\begin{array}{c @{\mbox{ for }} c}
    j_n(k_0r')h_n^{(1)}(k_0r) & r>r' \\
    j_n(k_0r)h_n^{(1)}(k_0r') & r\leq r' \\
\end{array}
\right.
,\label{eq:expexp}
\end{eqnarray}
where $h_n^{(1)}(z)$ are the spherical Hankel functions. We then introduce $f_n(r)$ such that 
\begin{equation}
\int \mbox{d}\mathbf{r}'\rho(r')\frac{\exp(i|\mathbf{r}-\mathbf{r}'|)}{k_0|\mathbf{r}-\mathbf{r}'|}u_n(r')Y_{n0}(\hat{r}')=f_n(r)Y_{n0}(\hat{r}).\label{eq:fnker}
\end{equation}
so that, assuming $\rho(r)=0$ for $r>R$
\begin{equation}
f_n(R)= 4\pi i h_n^{(1)}(k_0R)\int_0^R \rho(r') r^{\prime 2}u_n(r')j_n(k_0r') \mbox{d} r'.\label{eq:fnR}
\end{equation}
Then, the projection of Eq.(\ref{eq:gen}) on mode $n$ leads to
\begin{eqnarray}
j_n(k_0r)&=&(2\delta+i)\beta_n u_n(r)+\beta_n f_n(r).
\end{eqnarray}
Using Eq.(\ref{eq:fnR}) and defining
\begin{equation}
\lambda_n=4\pi\int_0^R \rho(r)r^{2}u_n(r)j_n(k_0r) \mbox{d} r
\end{equation}
we obtain
\begin{equation}
\beta_n=\frac{j_n(k_0R)}{(2\delta+i)u_n(R)+i\lambda_n h_n^{(1)}(k_0R)}.
\end{equation}
Thus, the exact solution to the scattering problem is obtained under the form of infinite series over the spherical modes,
as is reminiscent of Mie's solution~\cite{Mie1908,Bohren}.
In the latter approach, the coefficients of scattering for each mode are determined by using the continuity of the tangential
electric and magnetic fields, and of their derivatives, at the cloud boundaries. In our problem, we considered a scalar electric field only,
and assuming it is orthoradial, it reads~\cite{Bachelard11}
\begin{equation}
E_s(\mathbf{r})=-\frac{dk_0^2}{4\pi\epsilon_0} \int \mbox{d}\mathbf{r}' \rho(\mathbf{r}')\frac{\exp(ik_0|\mathbf{r}-\mathbf{r}'|)}{|\mathbf{r}-\mathbf{r}'|}\beta(\mathbf{r}').
\label{eq:Es}
\end{equation}
Now, the electric field  expresses in two different ways inside ($r<R$) and outside ($r\geq R$) the cloud
\begin{eqnarray}
E_s^{(out)}(\mathbf{r})&=&\frac{E_0}{2i}\sum_{n=0}^\infty (2n+1)i^n  \beta_n\lambda_n h_n^{(1)}(k_0r) P_{n}(\cos\theta)\label{eq:Eout}
\\ E_s^{(in)}(\mathbf{r})&=&\frac{E_0}{2i}\sum_{n=0}^\infty (2n+1)i^n  \beta_n P_{n}(\cos\theta)4\pi\nonumber
\\ &&\times \Bigg[\left(\int_0^r \mbox{d} r' r^{\prime 2}u_n(r')j_n(k_0r')\right) h_n^{(1)}(k_0r)\nonumber
\\ &&+ \left(\int_r^R \mbox{d} r' r^{\prime 2}u_n(r')h_n^{(1)}(k_0r')\right) j_n(k_0r)\Bigg]\label{eq:Ein}.
\end{eqnarray}
A straightforward calculation shows that $E_s(\mathbf{r})$ and $\partial E_s(\mathbf{r})/\partial r$ as given by eqs.(\ref{eq:Eout}) and (\ref{eq:Ein}) are continuous
at the cloud boundary $r=R$.
This allows to conclude that our solution is the same as the one proposed by Mie, although we did not make a direct use of continuity hypotheses for the electromagnetic fields. Let us also remark that our solution holds for {\it any} radial solution $u_n(r)$ of the Helmholtz equation, i.e. it applies to any spherical cloud with finite boundary at $r=R$.

\section{Mie resonances in nonuniform resonant media\label{sec:nonunif}}

The treatment of scattering in nonuniform media is difficult due to the lack of explicit solutions to the wave equation in that case, so
numerical approaches are the most common to tackle with this problem~\cite{Bohren}.
We shall here treat the case of homogeneous samples, and of those with a quadratic dependence of the atomic density over the radius.

\begin{figure}
\centering
\begin{tabular}{c}
\epsfig{file=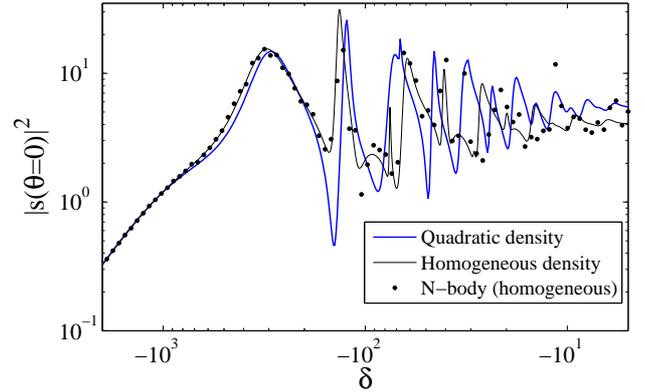,width=9cm}
\end{tabular}
\caption{Intensity scattered by the sample along the illuminating axis for homogeneous (thin black line) and quadratic (thick blue line) continuous densities, and for a $N$-atom homogeneous cloud with random atomic distribution (black dots). Simulations realized for $N=2.10^3$ and $\sigma_r=k_0\sqrt{\langle r^2 \rangle}=1$ (and thus $\sigma_r=\sqrt{3/5}k_0R$ for an homogeneous density, and $\sigma_r=\sqrt{3/7}k_0R$ for a quadratic one).\label{fig:Intensity}}
\end{figure}

In the first case, the atomic density of a cloud of radius $R$ is $\rho_0=3N/4\pi R^3$, whereas its refractive index is constant
$m_0=\sqrt{1-3N/[(k_0 R)^3(2\delta+i)]}$ and the solution of the wave equation inside the cloud is simply
$u_n(r)=j_n(k_0m_0 r)$.  Thanks to properties of the Bessel functions~\cite{Grad}, $\lambda_n$ can be calculated explicitely as
\begin{eqnarray}
\lambda_n&=&(2\delta+i) (k_0R)^2[m_0 j_{n-1}(k_0m_0R)j_n(k_0R)\nonumber\\
&&- j_{n-1}(k_0R)j_n(k_0m_0R)]\label{eq:lnhomo}
\end{eqnarray}
As for clouds of size $R$, with a parabolic density $\rho(r)=(5N/2V)[1-r^2/R^2]$ and volume $V=4\pi R^3/3$, their spatially-dependent refraction index reads
\begin{equation}
m_q^2(r)=m_c^2+\gamma^2 r^2,
\end{equation}
with $m_c=\sqrt{1- (15/2)N/[(k_0R)^3(2\delta+i)]}$ the index in the core of the sample, and $\gamma^2=(15/2)N/[(k_0R)^5(2\delta+i)]$.
Remark that when $k_0 R\to\infty$, but at constant central density $N/V$ and at fixed $r$, Eq.(\ref{eq:radeq}) tends toward the equation for the spherical Bessel
function $j_n(k_0 m_c r)$: the homogeneous medium limit is then recovered, as expected since the center of the cloud is then locally homogeneous.
Using the substitution~\cite{martin} $u_n(r)=r^{-3/2}w_n(x)$, with $x=\gamma r^2/2$, we obtain a Coulomb wave equation for $w_n(x)$, well known in nuclear physics~\cite{nuclear}
\begin{equation}
w_n''(x)+\left[1+\frac{m_c^2}{2\gamma x}-\left(\frac{n(n+1)}{4}-\frac{3}{16}\right)\frac{1}{x^2}\right]w_n(x)=0.
\end{equation}
Its solutions are the so-called Coulomb wave functions~\cite{abramowitz} $F_{n/2-1/4}(-m_c^2/4\gamma,x)$. The irregular Coulomb wave function $G$ is discarded because of its $1/r$ divergence at the origin, just as the spherical Hankel function is discarded to describe the field inside homogeneous media. Thus, we get the following solution for the radial component of the field
\begin{equation}
u_n(r)=\frac{1}{r^{3/2}}F_{\frac{n}{2}-\frac{1}{4}}\left(-\frac{m_c^2}{4\gamma},\frac{\gamma r^2}{2}\right).\label{eq:radcomp}
\end{equation}
Let us introduce the structure factor $s(\mathbf{k})=s(\theta,\varphi)$ for a direction $\mathbf{k}=k_0(\sin\theta\cos\varphi,\sin\theta\sin\varphi,\cos\theta)$, defined as
\begin{equation}
s(\mathbf{k})=\frac{1}{N}\int\mbox{d}\mathbf{r}' \rho(\mathbf{r}')\tilde\beta(\mathbf{r}')e^{-i\mathbf{k}\cdot\mathbf{r}'}.
\end{equation}
After integration over space, it turns into
\begin{equation}
s(\mathbf{k})=\sum_{n=0}^\infty (2n+1)\beta_n \lambda_n P_{n}(\cos\theta),\label{eq:sk}
\end{equation}
where we have made use of
\begin{eqnarray}
\int_0^{2\pi} \mbox{d}\varphi' &&\sin\varphi' e^{-i(a\cos\varphi'+b\sin\varphi')}=2\pi J_0\left(\sqrt{a^2+b^2}\right)\nonumber
\\ \int_{-1}^1\mbox{d} y &&J_0(r'\sin\theta\sqrt{1-y^2})P_n(y)e^{-iyr'\cos\theta}\mbox{d} y\nonumber
\\ &&=2i^{-n}j_n(r')P_n(\cos\theta).
\end{eqnarray}
Then, in the far-field limit ($r\gg R$), the scattered intensity simply reads
\begin{equation}
I_{scat}(r,\theta,\varphi)=\frac{E_0^2}{4(k_0r)^2}|s(\theta,\varphi)|^2.
\end{equation}
Using the analytic expressions obtained for the homogeneous and the parabolic density, we now turn to studying the resonances in Mie scattering for atomic clouds. Simulations of the intensity scattered along the illuminating axis reveal that homogeneous samples exhibit some spiky irregular structures, whereas inhomogeneous (parabolic) samples are characterized by more regular oscillations (see Fig.\ref{fig:Intensity}). This suggests not only  that the resonances do not only originate in the sharp boundaries of homogeneous obstacles, but on the contrary that they may even be easier to observe in media with smooth densities. Actually, the numerous spikes in homogeneous media can be interpreted as ``whispering gallery modes''~\cite{Rayleigh45,Nussenzveig92,Oraevsky02}, that propagate along the surface of the cloud, within a local minimum of the effective radial potential created by the jump of the medium's index. On the other hand, assimilating the cloud to a cavity, one can interpret the regular oscillations as cavity modes which survive despite the smoothening of the boundaries of the potential. Thus, the smoothening of the index leads to 
a modification of the local minimum for the effective potential which supports the whispering gallery modes~\cite{Nussenzveig92}, so that the latter may be attenuated or broadened (see Fig.\ref{fig:Intensity}), or even disappear (see Fig.\ref{fig:ResN}).

Finally, the simulations confirm that the theory for continuous media is in good agreements with $N$-body samples (see Fig.\ref{fig:Intensity}), although discrepancies appear for smaller values of detuning, and thus larger refractive index. However, simulations where $N$ is tuned reveal that this effect is due to the small number of particles that one can consider numerically: since finding the stationary solution of (\ref{eqbetaj}) requires the inversion of a $N\times N$ matrix problem, simulations are typically limited to systems of  at most $N\sim 2.10^3$ particles. Thus, the finite-$N$ effects observed in Fig.\ref{fig:Intensity} are expected to vanish for larger number of particles that one can typically find in a cold atomic cloud (that is $N=10^5-10^7$). More specifically, one can show that $N/\sigma_r^2\gg 1$ is required to neglect granularity~\cite{Bienaime10,Bachelard11}. This highlights the relevance of the continuous approach to model the scattering process of large-$N$ clouds, on which we shall now focus.

\section{Resonances in Mie scattering for cold atomic clouds\label{sec:mieresatoclo}}

\begin{figure}
\centering
\begin{tabular}{c}
\epsfig{file=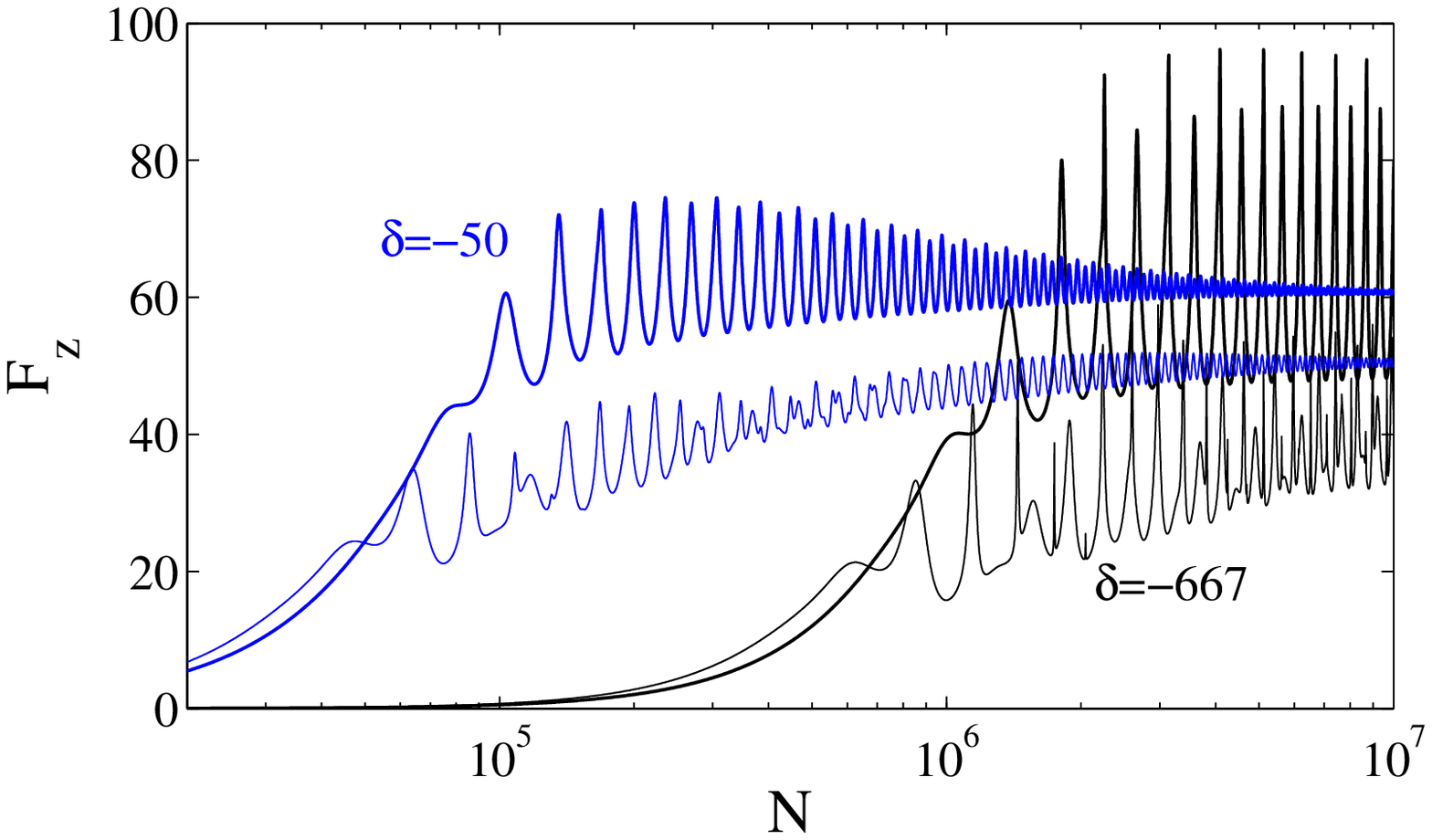,width=9cm}
\\ \epsfig{file=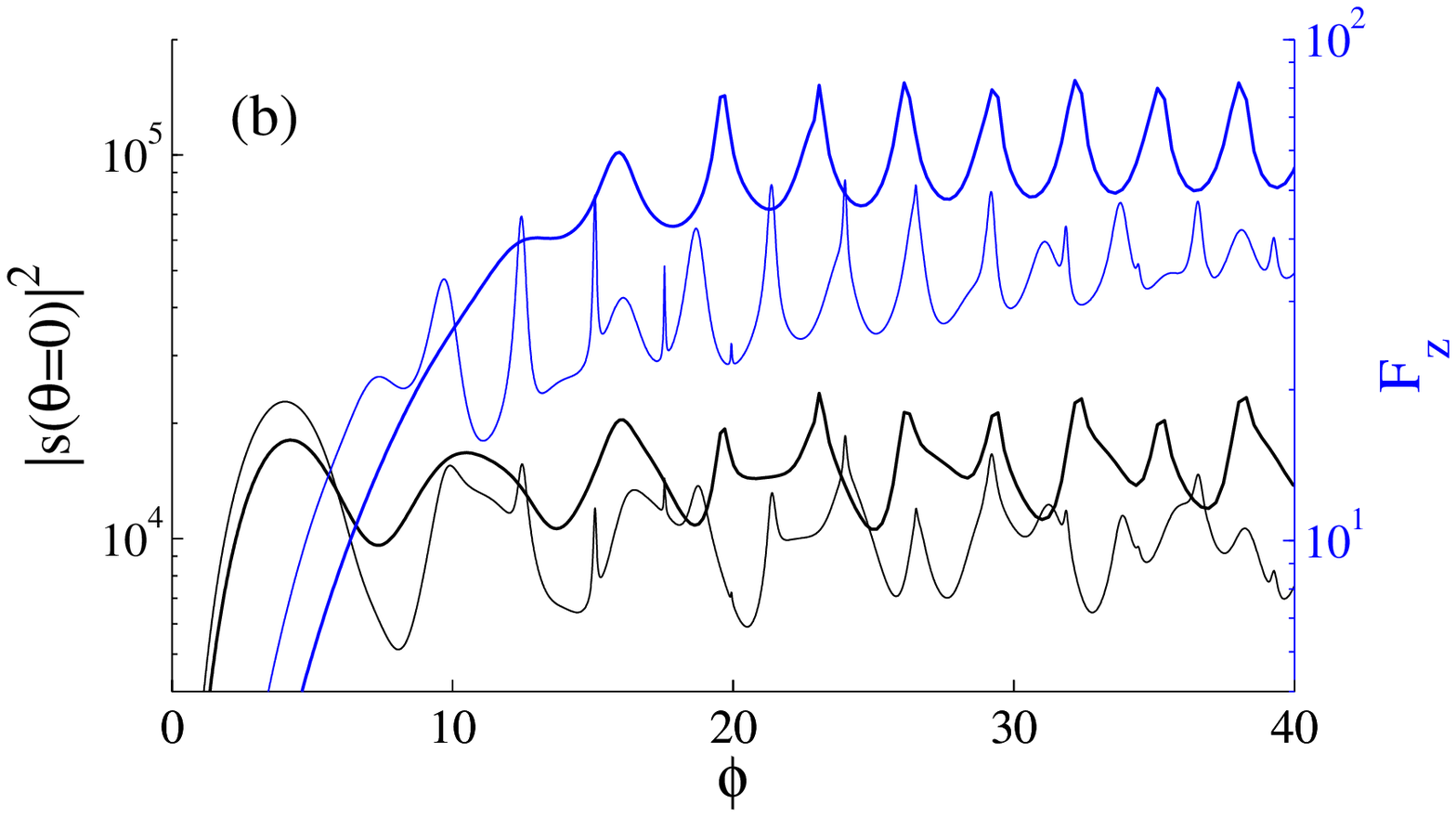,width=9cm}
\end{tabular}
\caption{(a) Radiation pressure force as a function of the number of atoms $N$ in the sample and for different detuning, namely $\delta=-667$ (black lines) and  $\delta=-50$ (blue lines). (b) Radiation pressure force as a function of the phase-shift $\phi$ for $\delta=-667$ (blue curves) and square of the structure factor $|s(\theta=0)|^2$ (black curves), proportional to the far-field scattered light in the forward direction. For both figures, the thin curves correspond to homogeneous samples, the thick ones to quadratic densities. Simulations realized for $\sigma_r=10$.\label{fig:ResN}}
\end{figure}

One of the effects of the light scattering by a cloud of cold atoms is the displacement of its center-of-mass (e.g. measured by time-of-flight imaging techniques~\cite{Bienaime10,Bender10}) due to the radiation pressure
force, which can be interpreted microscopically by a sequence of
absorption and spontaneous emission cycles. More specifically, the atom
absorbs a laser photon with $\mathbf{k}=k_0\hat{z}$ and re-emits it in a direction $\mathbf{k}=k_0(\sin\theta\cos\varphi,\sin\theta\sin\varphi,\cos\theta)$. The resulting radiation pressure force reads~\cite{Bachelard11}
\begin{equation}
F_z=-\hbar k_0\frac{\Omega_0^2}{\Gamma} \mbox{Im}(s(\mathbf{k}_0))-\frac{\hbar k_0 \Omega_0^2N}{8\pi\Gamma}\int \mbox{d}\mathbf{k}\cos\theta|s(\mathbf{k})|^2\label{eq:Fz}
\end{equation}
where $\Omega_0=dE_0/\hbar$ is the Rabi frequency of the incident field. In the framework of Mie scattering, the absorption part of the force reads
\begin{equation}
F_z^{(a)}=-\hbar k_0\frac{\Omega_0^2}{\Gamma N}\sum_{n=0}^\infty (2n+1)\mbox{Im}(\lambda_n\beta_n ).
\end{equation}
As for the emission part, using the orthogonality of Legendre polynomials, as well as the relation
\begin{equation}
\int_{-1}^1 x P_{n}(x)P_{n'}(x)\mbox{d} x=2\frac{\left(n\delta_{n,n'+1}+n'\delta_{n',n+1}\right)}{(2n+1)(2n'+1)},
\end{equation}
it reads
\begin{equation}
F_z^{(e)}=-\hbar k_0\frac{2\Omega_0^2}{\Gamma N}\sum_{n=0}^\infty (n+1)\mbox{Re}
( \lambda_n\lambda_{n+1}^*\beta_n \beta_{n+1}^*).\label{eq:fze0}
\end{equation}
These analytical expressions then allow to investigate numerically the presence of resonances in Mie theory in large-$N$ samples~\cite{Bienaime10,Bux10}, such as those produced in T\"ubingen~\cite{Bender10}, where clouds of size $\sigma_r=k_0\sqrt{\langle r^2 \rangle}\approx 10$ and $N\sim 10^6$ were studied. The simulations reveal that the radiation pressure force actually exhibits resonances when the number of atoms $N$ is tuned, at fixed detuning $\delta$ (see Fig.\ref{fig:ResN}a). Their contrast is significantly higher at (negative) larger values of $\delta$ yet they require larger number of atoms. Plotting the force oscillations as a function of the phase shift of on-axis light rays 
\begin{equation}
\phi=k_0\int \left[\mbox{Re}(m(0,0,z))-1\right]\mbox{d} z
\end{equation}
 makes clear that $\phi$ is the relevant variable to characterize the resonances in clouds with quadratic densities, i.e. cavity modes. Note that since the absorption and emission forces compensate at small $\phi$, Mie oscillations actually appear in the radiative force at values of $\phi$ higher than for the scattered light (typically $4\pi$).
The period of these oscillations can be qualitatively retrieved using the formulae for the homogeneous media (\ref{eq:lnhomo}). If one considers a large cloud $k_0R\gg 1$, then $j_n(x)\approx \sin(x-n\pi/2)/x$ (for $x\gg1$)
leads to the approximate expression
\begin{eqnarray}
\lambda_n&\approx& \frac{2\delta+i}{m_0}\Big[\frac{m_0+1}{2}\sin((m_0-1)k_0R)\label{eq:lnapp}
\\ &&+ \frac{m_0-1}{2}\sin((m_0+1)k_0R-n\pi)\Big],\ n^2<k_0R.\nonumber
\end{eqnarray}
This explains why the radiation pressure force will oscillate with the value of the phase-shift ($\phi=2\mbox{Re}(m_0-1)k_0R$ for homogeneous clouds, and $(4/3)\mbox{Re}(m_0-1)k_0R$ for samples with quadratic densities). Note that in the case of a dilute cloud such that $|m_0-1|\ll 1$ (yet $|m_0-1|k_0R$ can be large), one has $\phi\approx -(2/3) b_0\delta/(4\delta^2+1)$, which highlights the role of the optical thickness~\cite{Courteille10}
\begin{equation}
b_0=\frac{6\pi}{k_0^2}\int \rho(0,0,z)\mbox{d} z
\end{equation}
instead of the atomic density in the collective scattering problem~\cite{Bienaime10}. Since optical resonances will emerge for $N/\delta\sigma_r^2\sim 1$, the conditions for the observation reads $N>\delta\sigma_r^2$. Finally, remark that when the spatial density $N/\sigma_r^3$ becomes significant, other effects such as near field dipole dipole interactions have to be accounted for~\cite{Friedberg73,Sokolov09}.

\begin{figure}
\centering
\begin{tabular}{c}
\epsfig{file=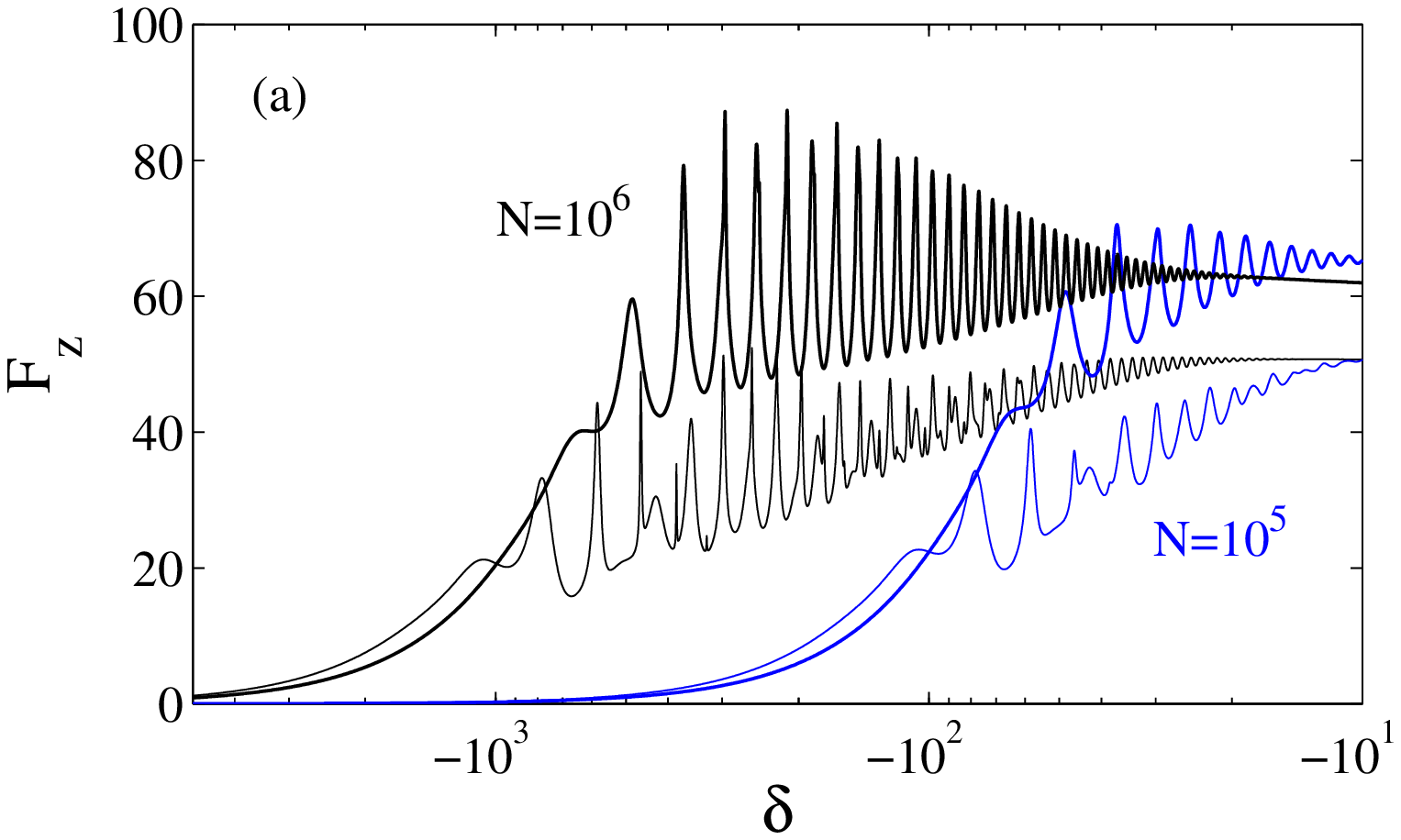,width=9cm}
\\ \epsfig{file=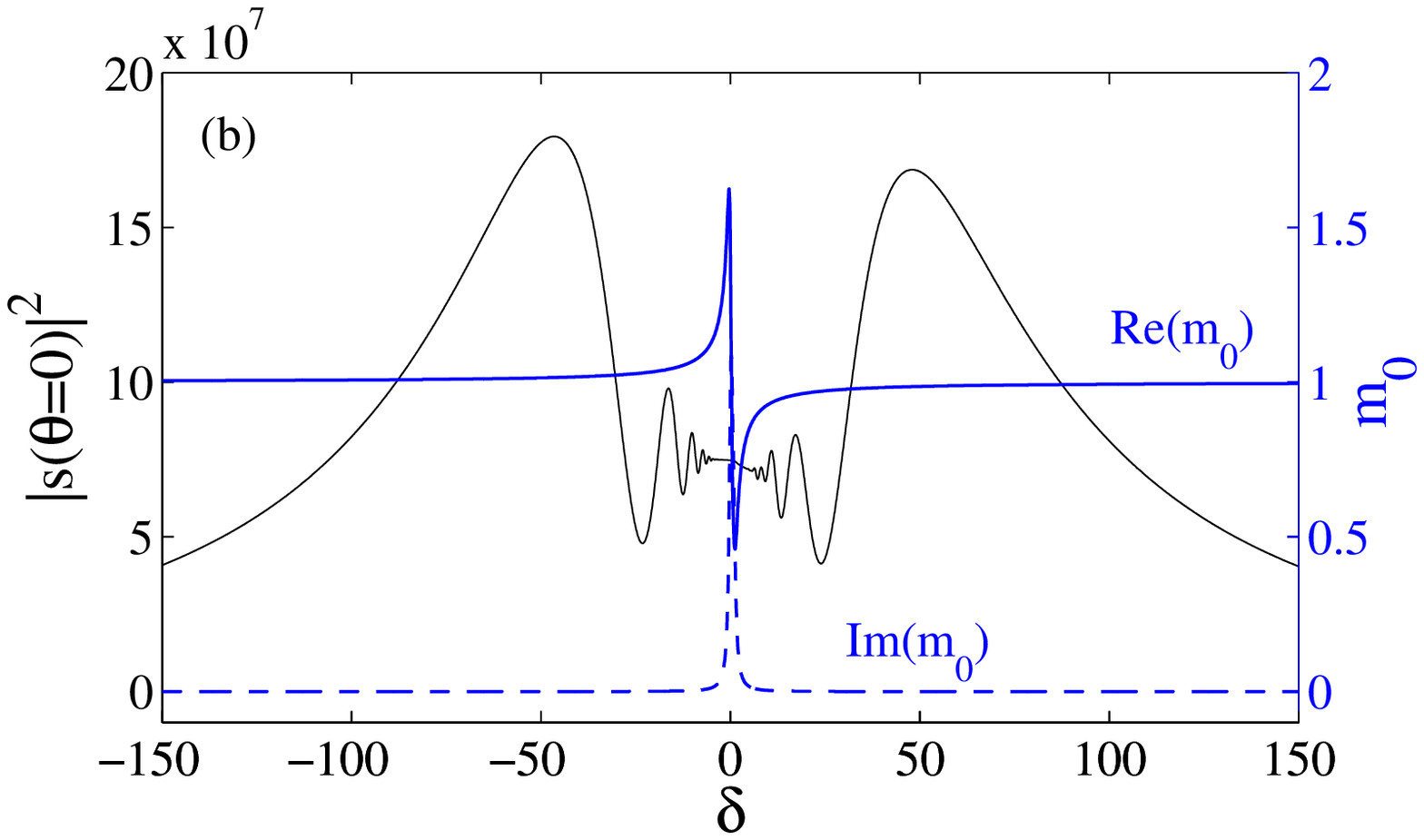,width=9cm}
\end{tabular}
\caption{(a) Radiation pressure force for a $\sigma_r=10$ cloud as a function of the detuning $\delta$ and for different numbers of particles, namely $N=10^6$ (black line) and  $N=10^5$ (blue line). The thin curves correspond to homogeneous samples, the thick ones to quadratic densities. (b) Intensity scattered along the illuminating axis $|s(\theta=0)|^2$ (black thin curve) as a function of the detuning, as well as real (plain blue line) and imaginary (dash-dotted blue line) parts of refractive index $m_0$ of the cloud, for a cloud of size $\sigma_r=100$ and homogeneously filled with $N=2.10^6$ particles.\label{fig:ResDelta}}
\end{figure}

Here, due to the special role of the wavelength of the illuminating laser on resonant atoms, the detuning $\delta$ can also be used as a control parameter to generate Mie oscillations: at odds from dielectric droplets experiments where it changes the effective size of the cloud $k_{\mbox{laser}}R$, it here modifies only the index of the cloud. Fig.\ref{fig:ResDelta}a depicts this phenomenon when the incident wavelength is tuned, and one recovers Mie oscillations in that case as well. As discussed previously, clouds with smooth (quadratic) densities exhibit more regular and more contrasted oscillations. Remark that using $\delta$ as a control parameter, the curves $F_z(\phi)$ or $I_{scat}[\theta=0](\phi)$ are extremely similar to those where $N$ is tuned (see Fig.\ref{fig:ResN}b), since only the refractive index is important, and the phase-shift $\phi$ captures this information.

Finally, we conclude by pointing out that the resonances in Mie theory can be observed as well at positive detunings: As can be observed in Fig.\ref{fig:ResDelta}b, the scattered intensity is rather symmetric in $\delta$, which can for example be understood by the fact that the phase-shift $\phi\approx -(2/3) b_0\delta/(4\delta^2+1)$ is antisymmetric in $\delta$ (yet the imaginary part of the refractive index is not, thus the difference in amplitudes). The symmetry decreases with $\delta$, when the refractive index deviates significantly from unity, yet model (\ref{eqbetaj}) then loses its validity, and a more exhaustive model is required to study the light scattering~\cite{Friedberg73}.

\section{Conclusion}

We studied resonances in Mie theory in resonant media with parabolic densities, a case most appropriate for cold atomic clouds. 
It was shown that the Mie oscillations were more regular in the case of smooth (parabolic) densities, where only cavity modes survive, and that they can be observed as well in the scattered intensity as in the radiation pressure force, even though for the lowest phase shifts the radiation pressure force is a less sensitive measurement.

The parameters of Figs.\ref{fig:ResN}a and \ref{fig:ResDelta}a correspond to the experimental case of~\cite{Bender10}, and it suggests that the resonances can be detected in the available range of parameters. Note that the main difference with this experiment is that the cloud was then cigar-shaped, and the density assumed to be Gaussian; yet parabolic densities are definitely more realistic than homogeneous ones to describe the Gaussian case.

Furthermore, the contrast of the resonances is connected to the imaginary term of the cloud's index (\ref{eq:nr}), as discussed in~\cite{Prasad11}. More specifically, the smaller the single-atom decay term is (compared to the detuning), the larger the contrast. Hence, large-$N$ and far-detuned configurations will generate sharper resonances, for they minimize the single-atom decay contribution.

A more thorough treatment would of course require considering vectorial fields~\cite{Prasad11}. Nevertheless, simulations reveal that the scalar approach provides an accurate description of resonances in Mie scattering for large and dilute atomic clouds.

Finally, we note that tuning the illuminating laser toward the blue ($\delta>0$) can also allow for the observation of the resonances. However, when $4\pi\rho/(2\delta k_0^3)>1$, the sample becomes strongly dissipative since $\mbox{Re}(m^2)$ becomes negative (see Eq.(\ref{eq:nr})): radiation trapping will then emerge~\cite{Holdstein47}, that will damp the scattering process and eclipse the resonances. The occurence of this phenomenon in cold atomic clouds will be the object of future works.


\begin{thebibliography}{99}
\bibitem{note:Prasad} Notice that the imaginary term in Eq.(\ref{eq:nr}), taking into account absorption and so the incoherent damping of the
excitation, was not considered in ref.\cite{Prasad11} and only phenomenologically included in the last section of that paper.
In the present analysis it is included in the model from the beginning.
\end{thebibliography}
\end{document}